\documentclass[aps,prb,amsmath,amssymb,twocolumn,10pt,footinbib]{revtex4-1}
\usepackage{graphicx}
\usepackage{hyperref}

\begin{document}
\title{Non-magnetic defects in the bulk of two-dimensional topological insulators}

\author{Vladimir~A.~Sablikov and Aleksei~A.~Sukhanov}

\affiliation{V.A. Kotel’nikov Institute of Radio Engineering and Electronics, Russian Academy of Sciences,
Fryazino, Moscow District, 141190, Russia}

\begin{abstract}
We found that non-magnetic defects in two-dimensional topological insulators induce bound states of two kinds for each spin orientation: electron- and hole-like states. Depending on the sign of the defect potential these states can be also of two kinds with different distribution of the electron density. The density has a maximum or minimum in the center. A surprising effect caused by the topological order is a singular dependence of the bound-state energy on the defect potential.
\end{abstract}
\maketitle

\section{Introduction}
\label{Intro}

Topological insulators (TIs) are of great interest due to amazing physical properties of the electronic system caused by the presence of topological order~\cite{Hasan,Franz_Molenkamp}. The most attention is paid to the topologically protected surface and edge states that exist at the boundaries of the topologically non-equivalent crystals because of strong spin-orbit interaction. They are protected by the time-reversal symmetry against weak non-magnetic impurities or disorders. Owing to the presence of the protected states the TIs are expected to have good surface conductivity and large bulk resistivity. However the TI samples available nowadays are always poorly insulating in the bulk, owing to a large amount of structure imperfections (defects)~\cite{Franz_Molenkamp}. The questions of what is the electronic and topological structure of the defects and how they affect the properties of the TIs attract increasing interest in recent years~\cite{Teo,Lee}.

In this Letter we focus on the non-magnetic defects. The investigations of the electronic structure of the defects on the surface of three dimensional (3D) TIs reveal a rich interference pattern and non-trivial spin texture arising around the defects due to quasiparticle scattering~\cite{Zhou,Guo,Wang,Biswas}. Non-magnetic impurities and vacancies in surface, subsurface, and bulk positions in 3D TIs modify the energy spectrum of the surface states: they may disrupt the Dirac cone and create in-gap resonances~\cite{Black-Schaffer1,Black-Schaffer2}. The theoretical studies of an Anderson impurity in the bulk of TIs have shown that a Kondo resonant peak appears simultaneously with an in-gap bound state in the case where the band-dispersion has a Mexican-hat form~\cite{Lu,Kuzmenko}. 
	
The formation of in-gap bound states in two-dimensio\-nal (2D) TIs was demonstrated by considering a hole in which the wave function is zero~\cite{Shan}. The bound states have the same origin as the edge states circulating around the hole with quantized angular momentum. The edge states appear in pairs propagating in opposite directions with opposite spins. A point defect in the crystal essentially differs from the blowhole since the wave function is not zero there, but should be found taking into account the potential of the defect. This is a rather complicated problem because of multiband structure of the Hamiltonian. The bound states induced by the Gaussian potential were numerically investigated by Sh.-Q. Shen et al~\cite{J_Lu,Shen} for a number of material parameters. It was found that the bound states appear under certain conditions but, no general conclusions were done about their spectra, electronic structure and the conditions under which they exist.

We have found a way to solve this problem thoroughly in the physically interesting case where the defect potential is strongly enough localized. In this paper we study analytically the bound states induced by a non-magnetic defect in 2D TIs and clarify their general properties: the classification of the states, the conditions under which they exist, their electronic and current structure. Particularly, we find that two kinds of the states exist near the defect for each spin direction, in contrast to the edge states on a smooth boundary and the states on the hole. 

\section{The model}\label{model}
Our consideration is based on the model of the 2D TIs proposed by Bernevig, Hughes and Zhang for HgTe/CdTe quantum wells~\cite{BHZ}. The Hamiltonian has the form
\begin{equation}
 H=
\begin{pmatrix}
 h(\mathbf{k}) & 0\\
 0 & h^*(-\mathbf{k})
\end{pmatrix}\,,
\label{HH}
\end{equation} 
where $\mathbf k$ is momentum operator and
\begin{equation}
h(\mathbf{k})=
\begin{pmatrix}
 M\!-\!(B\!+\!D)k^2 & A(k_x+ik_y)\\
 A(k_x-ik_y) & -M\!+\!(B\!-\!D)k^2
\end{pmatrix}\,,
\label{H}
\end{equation} 
where $M$, $A$, $B$ and $D$ are model parameters. In the topological phase $MB>0$. In the case of HgTe/CdTe wells $M, B, D<0$, $A>0$. The basis set of wave functions is composed of the electron and heavy-hole sub-band states with opposite spins: $\{|E_1\uparrow\rangle,|H_1\uparrow\rangle,|E_1\downarrow\rangle,|H_1\downarrow\rangle\}$. 

The defect is described by the potential $V(\mathbf r)$ localized in a small region near $r=0$. Since the defect is non-magnetic, the Hamiltonian (\ref{HH}) is separated into spin blocks for each of which the Schr\"odinger equation has the form
\begin{equation}
\left[E\,\mathbf{I}_2-h(\mathbf{k})\right]\Psi(\mathbf r)=\mathbf{I}_2V(\mathbf{r})\Psi(\mathbf r)\,,
\end{equation} 
where $\mathbf{I}_2$ is a $2\times2$ unit matrix, $\Psi(\mathbf{r})$ is a two-component spinor $(\psi_1(\mathbf{r}),\psi_2(\mathbf{r}))^T$. The wave functions are supposed to vanish at infinity.

In terms of the momentum-space wave functions 
\begin{equation}
 \Phi(\mathbf k)=\iint d^2\mathbf{r}\,\Psi(\mathbf r)e^{-i\mathbf{kr}}\, 
\end{equation} 
the Schr\"odinger equation reads
\begin{equation}
 [E\,\mathbf{I}_2-h(\mathbf{k})]\Phi(\mathbf{k})=\mathbf{I}_2\iint d^2\mathbf{r}\,V(\mathbf{r})\Psi(\mathbf{r})e^{-i\mathbf{kr}}.
 \label{Eq_1}
\end{equation}

The key point is to treat the integral in the right-hand side. One can naively suppose that $V(\mathbf{r})$ is a $\delta$ function, but in this case the problem has no solution. This is a well-known property of 2D and 3D systems~\cite{Frank,Jackiw} which was recently confirmed for TIs~\cite{Lu}. However, in reality the potential is not strictly $\delta$ function. One of effective ways allowing one to overcome this difficulty is to introduce the cut-off into the integration over $\mathbf{k}$ at large $k$ and subsequently regularize the problem in the momentum space~\cite{Jackiw}. Alternatively one can simplify the integration using the fact that $V(r)$ has a sharp maximum in the point $r=0$. In our case both approaches give close results, but the approximation of $V(r)$ by a sharp regular function is more justified from the standpoint of the experimental realization of 2D TIs in quantum well heterostructures, especially in the case of charged defects. 

If the potential $V(\mathbf{r})$ is localized in a region small compared to characteristic lengths of the wave-function variation, the integral in the right-hand side of Eq.~(\ref{Eq_1}) can be simplified by expanding $\Psi(r)$ in $r$. Such an expansion is justified when $|\Psi''(0)/\Psi(0)|\Lambda^{-2}\ll 1$, where $\Lambda^{-1}$ is the size of the potential localization region. 

Using this simplification and supposing that the potential is axially symmetric, we arrive at the following equations for the Fourier components $\Phi_1(k)$ and $\Phi_2(k)$ of the wave-function spinor:
\begin{equation}\left\{
 \begin{array}{rl}
  \![E\!-\!M\!+\!(B\!+\!D)k^2]\Phi_1\!-\!A(k_x\!+\!ik_y)\Phi_2&\!=\!2\pi\overline{\psi}_1V_k\\
 \!-A(k_x\!-\!ik_y)\Phi_1\!+\![E\!+\!M\!-\!(B\!-\!D)k^2]\Phi_2&\!=\!2\pi\overline{\psi}_2V_k,
 \end{array}\right.
 \label{Eq_2}
\end{equation}
where $\overline{\psi}_{1,2}\equiv\psi_{1,2}(0)$, $V_k$ is the Hankel transform of $V(r)$,
\begin{equation}
 V_k= \int_0^{\infty}\!drrJ_0(kr)V(r)\,.
\end{equation} 

Using Eqs~(\ref{Eq_2}) one can easily find the wave functions
\begin{equation}
\begin{split}
 \psi_{1,2}(\mathbf{r})\!=&\overline{\psi}_{1,2}\!\int_0^{\infty}\!dkk\frac{E\!+\!Dk^2\!\pm\!(M\!-\!Bk^2)}{\Delta(E,k)}V_kJ_0(kr)\\
 &+i\overline{\psi}_{2,1}e^{\pm i\varphi}\!\int_0^{\infty}\!dk\frac{Ak^2}{\Delta(E,k)}V_kJ_1(kr),
\end{split}
\label{psi1}
\end{equation} 
where $\varphi$ is the azimuthal angle of the vector $\mathbf{r}$ and $\Delta(E,k)=(E\!+\!Dk^2)^2\!-\!(M\!-\!Bk^2)^2\!-\!A^2k^2$. 
 
By writing these equations at $r=0$, where $\psi_{1,2}(r=0)=\overline{\psi}_{1,2}$, we obtain a set of equations which determine the eigenenergy and $\overline{\psi}_{1,2} $. Because $J_1(kr=0)=0$, the system is decoupled into two independent equations:
\begin{equation}
 \overline{\psi}_{1,2}\left[1-\int_0^{\infty}\!dkk\frac{E\!+\!Dk^2\!\pm\!(M\!-\!Bk^2)}{\Delta(E,k)}V_k\right]=0.
 \label{states}\\
\end{equation}
The equations show that there are two kinds of eigenstates:\\
i)~~the states in which $\overline{\psi}_1\ne 0$ and $\overline{\psi}_2=0$,\\
ii)~~the states in which $\overline{\psi}_1= 0$ and $\overline{\psi}_2\ne 0$.

Taking into account the arrangement of the wave functions in the basis set one can conditionally say that the first state is electron-like in the center, and second one is hole-like. Their eigenenergies are determined by zeros of the expression in the square brackets in Eq.~(\ref{states}),
where the upper and lower signs relate to the electron-like bound states $|e\rangle$ and hole-like bound states $|h\rangle$.

Eqs~(\ref{states}) clearly show that no bound states exist when the potential $V(r)$ is the $\delta$ function. In this case $V_k$=Const and the integrals diverge logarithmically showing that $\Psi(r)=0$. However, if the potential, even of zero radius, has a singularity more weak than the $\delta$ function, its Hankel transform $V_k$ decreases with $k$ and the integrals converge. Therefore bound states can exist. Physically realistic potential has of course a finite radius, $\Lambda^{-1}$. In this case Eqs~(\ref{states}) can be solved for specific $V(r)$.

\section{Results and discussion\label{Results}}
Below the detail results are presented for the Gaussian potential: $V(r)=v\Lambda^2/\pi\exp[-\Lambda^2r^2]$. Its Hankel transform is $V_k=v/(2\pi)\exp[-k^2/(4\Lambda^2)]$. With this $V_k$ the integrals in Eqs~(\ref{states}) are calculated analytically. 

For simplicity and without loss of generality, we put $D=0$. In this case the problem contains important parameters $a=A^2/(2MB)$ and $v/B$ that essentially determine the spectrum of the band states. If $0<a<1$, the band dispersion has a Mexican-hat form and the bulk energy gap is smaller than $2|M|$. When $a>1$ the dispersion is quadratic and the gap equals $2|M|$. At $a=1$ the band dispersion is flat in the vicinity of the point $k=0$.

The analysis of Eqs~(\ref{states}) shows that each equation has only one root at a given $v/B$, namely $E_e$ for the electron-like states and $E_h$ for the hole-like states. Both roots are connected by the relation $E_e(v/B)=-E_h(-v/B)$, so it is enough to consider one of them. We will consider $E_e$. It is interesting that each eigenenergy has a quite different dependence on $v/B$ for positive and negative $v/B$. This is because the bound states $|e\rangle$, which are formed at $v/B<0$ and $v/B>0$, have different electronic structure. Therefore we denote $E_{e1}=E_e(v/B<0)$ and $E_{e2}=E_e(v/B>0)$ and similarly for the hole-like states. 

\begin{figure}
\centerline{\includegraphics[width=1.\linewidth]{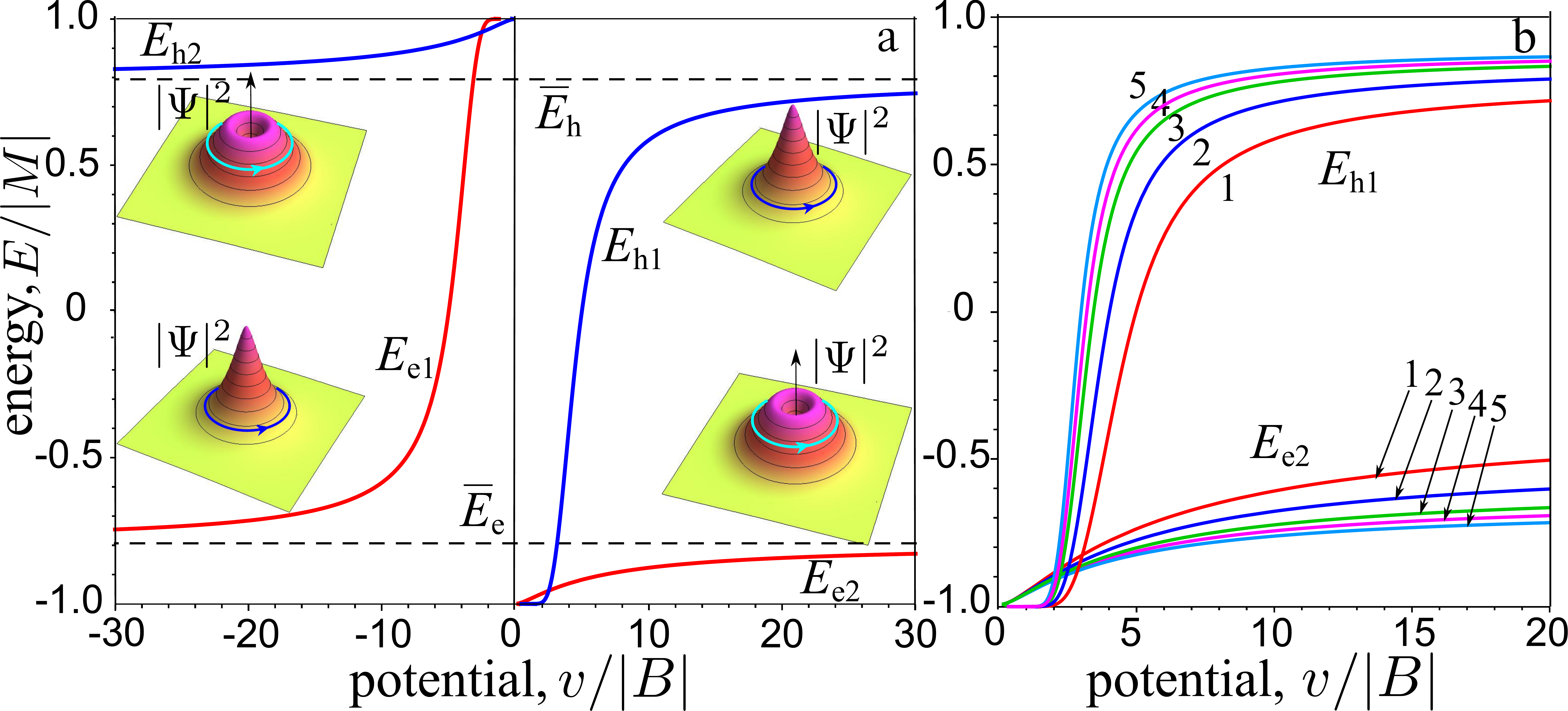}}
\caption{(a) The  energy of the electron- and hole-like bound states as a function of the defect potential. Lines $E_{e1}$, $E_{e2}$ and $E_{h1}$, $E_{h2}$ correspond to two kinds of the electron- and hole-like states (see text). Dashed lines show the limiting energies. The calculations were carried out for $a=1$, $\lambda=5$ and $B<0$. (b) The bound-state energies $E_{e2}$ and $E_{h1}$ versus the potential for a variety of $\lambda$. Lines 1, 2, 3, 4, 5 correspond to $\lambda=5, 7, 10, 15, 20$. For clarify the vertical scale of $E_{e2}$ lines is increased by 5 times.}
\label{fig_E-v}
\end{figure}

The dependence of the eigenenergies on $v/B$ is illustrated in Fig.~\ref{fig_E-v}a. This picture is qualitatively similar for all parameters, where the topological phase exists. 

If $v/B>0$, the electron-like bound state $|e1\rangle$ arises when $v/B$ exceeds a threshold value $v_{th}/B$ which depends on the potential-localization parameter $\lambda=\Lambda\sqrt{B/M}$ and the parameter $a$. The threshold value $v_{th}/B=0$ when $a<1$. For $a>1$, the following estimation is obtained
\begin{equation}
 v_{th}/|B|\approx 4\pi/\left\{\ln(4\lambda e^{-\gamma})-\ln[2(a-1)]\right\},
 %\frac{v_{th}}{B}\approx \frac{4\pi}{\ln(4\lambda e^{-\gamma})-\ln[2(a-1)]},
\label{v_th}
\end{equation} 
where $\gamma$ is the Euler constant. Here we have taken into account that the potential is strongly localized and $\lambda \gg 1$.

At the threshold the energy of the state $|e1\rangle$ lies at the top of the bulk gap. With increasing $v/B$ the energy $E_{e1}$ goes down to a limiting value $\overline{E}_e$, which is reached asymptotically. $\overline{E}_e$ also depends on the parameters $\lambda$ and $a$. Particularly for $a>1$, $\overline{E}_e$ is estimated as
\begin{equation}
 \overline{E}_e/|M|\approx -1+2^a(a-1)^a/\left(4\lambda e^{-\gamma}\right)^{(a-1)}.
 %\frac{\overline{E}_e}{|M|}\approx -1+\frac{2^a(a-1)^a}{\left(4\lambda e^{-\gamma}\right)^{(a-1)}}.
 \label{E_th}
\end{equation} 

When $v/B<0$, the bound state $|e2\rangle$ exists in the entire region $v/B<0$ and its energy increases from the bottom of the gap at $v/B=-0$ up to $\overline{E}_e$ at $v/B\to -\infty$. 

In the case where $0<a<1$, the energy gap is less than $2|M|$, but the bound-state energy depends on $v/B$ similarly so that the limiting energies always lie in the gap.

Energy of the bound states and their very existence depend not only on the defect potential, but also on the localization parameter $\lambda$, though this dependence is logarithmically weak. With increasing $\lambda$ the graphs of the bound-state energy dependence on $|v|$ are pressed to the edges of the gap and to the vertical line $v=0$, as Fig.~\ref{fig_E-v}b shows. It is worth noting that in the limit $\lambda\to \infty$ the bound states disappear since $v(r)$ becomes the $\delta$ function.

The presence of two kinds of states, $|e1\rangle$ and $|e2\rangle$, and the limiting energy dividing them is the most remarkable feature that exists only in the topological phase. Direct calculations for two-band model of normal insulator ($MB<0$) show that only one bound state exists at a given $v$: electron-like state for $v/B<0$ and hole-like state for $v/B>0$. It is evident that two kinds of states originate from the topological properties of the crystal. 

To clarify their nature we consider the electronic and current structure of $|h1\rangle$ and $|e2\rangle$ states at a given $v/B<0$ using the wave functions (\ref{psi1}).

\begin{figure}
\centerline{\includegraphics[width=1.\linewidth]{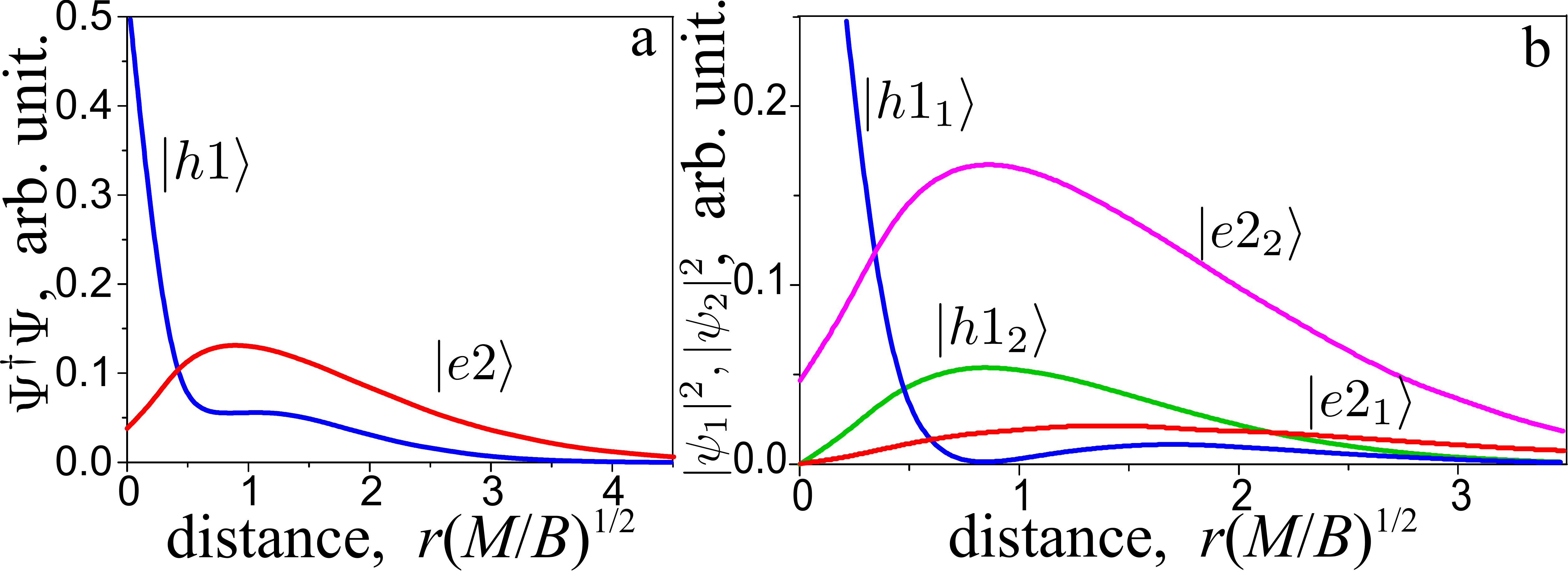}}
\caption{a) The radial distribution of the density $\Psi^{\dag}\Psi$ in the bound states $|h1\rangle$ and $|e2\rangle$. b) Densities of the spinor components. $|h1_{1,2}\rangle$ and $|e2_{1,2}\rangle$ denote the components $\psi_1$ and $\psi_2$ in the states $|h1\rangle$ and $|e2\rangle$. For convenience the scale of the $|e2\rangle$ densities is increased 5 times. The calculations were done for $v=4$, $a=1$ and $\lambda=6$.}
\label{fig_dens12}
\end{figure}

The radial distribution of the densities $\Psi^\dag(r)\Psi(r)$ in both states are shown in Fig.~\ref{fig_dens12} together with partial densities of the spinor components $|\psi_1(r)|^2$ and $|\psi_2(r)|^2$. It is seen that the distribution of the electron density in the states of the first and second kind differs qualitatively. In the first-kind states, $|e1\rangle$ and $|h1\rangle$, the density has a maximum in the center. With increasing the radius the density varies, generally speaking, non-monotonically. At high potentials $|v|$ it can reach a maximum at a certain distance from the center. In contrast, the density in the second-kind states, $|e2\rangle$ and $|h2\rangle$, has a minimum at the center, and reaches a maximum at some distance from it.

The current densities are calculated using the current operator, which is easily obtained from the continuity equation and the Hamiltonian (\ref{H}). In our case the angular component is only important:
\begin{multline}
 j_{\varphi}=-\frac{2}{\hbar}\mathrm{Im}\left[(B+D)\psi^*_1\frac{\partial\psi_1}{\partial \varphi}-(B-D)\psi^*_2\frac{\partial\psi_2}{\partial \varphi}\right]\\
 -\frac{2}{\hbar}A\,\mathrm{Im}\left[e^{i\varphi}\psi^*_1\psi_2\right].
\label{current}
\end{multline}
Here the non-trivial second term not containing the spatial derivatives is caused by the non-diagonal elements of the Hamiltonian~(\ref{H}), quite similarly to the electron systems with spin-orbit interaction~\cite{Sonin}.

Using the wave functions~(\ref{psi1}) we have found that in both states, $|h1\rangle$ and $|e2\rangle$ at a given $v/B<0$, the electron current flows around the defect in the same direction for a given spin that coincides with the direction of the edge currents near the smooth boundary. An essential difference with the edge states is that the bound state on the defect can be occupied by only one electron with a certain spin. The probability of finding a second electron on the defect is small because of electron-electron repulsion. In contrast, the edge state near a smooth boundary can be occupied by two counter-moving electrons with opposite spins so that the electron current is absent.

These properties of the bound states can be understood as follows. In normal insulators the only reason of the bound state formation is an attractive potential for electrons or holes. In TIs there is another reason caused by the boundary condition for the wave function. This mechanism is realized when there is a hole in the crystal~\cite{Shan}. If the defect potential is finite and localized in a small region, both above mechanisms work. The states of the first kind originate mostly from the potential attracting the electrons or holes. The second-kind states originate from the edge states. Of course, in both states there is a circulating current and the {electron density is not zero in the center.

The singularity of the dependence of the bound-state energies on the defect potential at the levels $\overline{E}_{e,h}$ can have an interesting consequence. Due to this feature, the bound-state energies very weakly depend on $v$ when $|v|\gg1$ and are located very close to the limiting energies. Since in reality the crystal contains a large number of various defects, their potentials are distributed in a wide range. The typical potentials $V(r)$ are of the order of 1-10~eV. The estimation of $|v/B|$ for Hg/CdTe well parameters gives $|v/B|\sim 3-30$. Therefore the bound-state energies of the different defects lie in narrow intervals near $\overline{E}_{e,h}$. Since this energy is close to the bands the spatial size of the bound states is very large, $\sim 10^{-5}$~cm. Hence, the bound states can overlap even at relatively small density of defects, $\sim 10^{10}$~cm$^{-2}$, and  form a hopping or collective state which can manifest itself in transport. 

\section{Summary\label{Summary}}
In conclusion, we have investigated the bound states induced by non-magnetic defects with short-ranged potential in the bulk of 2D TIs. We have found that  the electron- and hole-like states exist for each spin orientation. These states, in turn, can be of two kinds depending on the sign of the defect potential. The first-kind states originate mostly from the potential attracting electrons or holes. The states of the second kind arise from the edge states of TIs. In both states there is an electron current. The most remarkable feature of the bound-state spectrum is the singularity of the dependence of the bound-state energy on the defect potential due to which the bound states can overlap even at relatively small density of defects. 

%%%%%%%%%%%%%%%%%%%%%%%%%%%%%%%%%%%%%%%%%%%%%%%%%%%%%%%%%%%%%%%%%%%%%%%%%%%%%%%%%%%%%%%%%%%%%%%%%%%%%%%%%%%%%%%%%%%%%%%%%%%
\acknowledgments
This work was partially supported by Russian Foundation for Basic Research (project No~14-02-00237) and programs of Russian Academy of Sciences.

\end{document}